\documentclass{elsart}
\usepackage[all]{xy}
\usepackage{graphicx}
\journal{Physics Letters B}
\begin{document}

%
\begin{frontmatter}
\title{3-3-1 Models at Electroweak Scale}

\author[label1]{Alex G. Dias}
\author[label2]{J. C. Montero}
\author[label2]{V. Pleitez}
\address[label1]{Instituto de  F\'{\i}sica, Universidade
de S\~ao Paulo,\\ Caixa Postal 66.318, 05315-970, S\~ao Paulo-SP, Brazil}
\address[label2]{Instituto de F\'{\i}sica Te\'orica, Universidade
Estadual Paulista \\
Rua Pamplona, 145, 01405-900 S\~ao Paulo, SP - Brazil}

\begin{abstract}
We show that in 3-3-1 models there exist a natural relation among the
$SU(3)_L$ coupling constant $g$, the electroweak mixing angle $\theta_W$, the
mass of the $W$,  and one of the vacuum expectation values,
which implies that those models can be realized at low energy scales and, in
particular, even at the electroweak scale. So that, being that symmetries
realized in Nature, new  physics may be really just around the corner.
\end{abstract}

\begin{keyword}
3-3-1 model \sep neutral currents
\PACS 12.10.Dm \sep 12.10.Kt \sep 14.80.Mz
\end{keyword}

\end{frontmatter}

\date{\today}

\maketitle


\textit{Introduction}.---
Many of the extension of the  Standard Model (SM)  implies the exis\-ten\-ce
of at least one extra neutral vector boson, say $Z^\prime$, which
should have a mass of the order of few TeV in order to be consistent with
present phenomenology. This is the case, for instance, in
left-right models~\cite{lr}, any grand unified
theories with symmetries larger than $SU(5)$ as $SO(10)$ and
$E_6$~\cite{hewett}, li\-ttle Higgs scenarios~\cite{lh}, and models with
extra dimensions~\cite{kk}. This makes the search for extra neutral
gauge bosons one of the main goals of the
next collider experiments~\cite{lhcilc}. Usually, the interactions involving
$Z^\prime$  are parametrized [besides the pure kinetic
term] as~\cite{babu,pdg}
\begin{eqnarray}
{\mathcal  L}^{NC(Z^\prime)}&=&-
\frac{\sin\xi}{2}F^\prime_{\mu\nu}F^{\mu\nu}+M^2_{Z^\prime}Z^\prime_\mu
Z^{\prime\mu}+\delta M^2Z^\prime_\mu Z^{\mu}\nonumber \\ &-&
\frac{g}{2 c_W}
\sum_i\bar{\psi}_i\gamma^\mu(f^i_V-f^i_A\gamma^5)\psi_i Z^\prime_\mu,
\label{zgeral}
\end{eqnarray}
where $Z$, which is the would be neutral vector boson of the SM,
and $Z^\prime$ are not yet mass eigenstates, having a mixing defined by
the angle $\tan 2\phi=\delta M^2/(M^2_{Z^\prime}-M^2_Z)$; $c_W\equiv \cos \theta_W$
[and for future use $s_W\equiv \sin \theta_W$]. If $Z_1$ and
$Z_2$ denote the mass eigenstates, then in most of the models
$M_{Z_2} \gg M_{Z_1}\approx M_Z$. In this situation the  vector  and
axial-vector couplings, $g^{(SM)}_V$ and $g^{(SM)}_A$, respectively,
of the SM $Z$ boson with the known fermions are modified at tree
level  as follows:
\begin{equation}
g^i_V= g^{i(SM)}_Vc_\phi+ f^i_V s_\phi,\; g^i_A= g^{i(SM)}_Ac_\phi+
 f^i_As_\phi,
\label{shift}
\end{equation}
where $g^{i(SM)}_V\!\!=\!\!T^i_3-2Q_i s_W^2$ and
$g^{i(SM)}_A=T^i_3$, being $T^i_3=\pm1/2$ and $Q_i$ the electric
charge of the fermion $i$;  we have used the notation
$c_\phi(s_\phi)\!\!=\!\!\cos\phi(\sin\phi)$.  The coefficients
$f^i_{V,A}$ in Eq.~(\ref{shift}) are not in general the same for all
particles of the same electric charge, thus, $Z^\prime$ induces
flavor changing neutral currents (FCNC) which imply strong
constraints coming from experimental data such as $\Delta M_K$ and
other $\vert\Delta S\vert=2$ processes. These constraints imply a
small value for the mixing angle $\phi$ or, similarly, a large value
to the energy scale,  generically denoted by $\Lambda$, related with
the larger symmetry. If  $s_\phi=0$ is imposed such constraints
could be avoided, however in most of the models with $Z^\prime$ this
usually implies a fine tuning among $U(1)$ charges and vacuum
expectation values, that is far from being natural~\cite{carena}.

Here we will show that there are models in which, at the tree level,
it is possible that: i) there is no mixing between $Z$ and
$Z^\prime$, and the latter boson may have a mass even below the TeV
scale; ii) $\rho_0=1$ since $M_{Z_1}=M_Z$, and iii) the couplings of
$Z_1$ with fermions,  $g^i_{V,A}$,  being exactly those of the SM,
$g^{i(SM)}_{V,A}$, no matter how large is $\Lambda$. This is implied
not by a fine tuning but by a condition which can be verified
experimentally involving the parameters of  the model, $g$, $M_W$,
$s_W$ and one  vacuum expectation value (VEV).

\textit{The model}.--- The so called 3-3-1 models are interesting
extensions of the standard model~\cite{331,pt,mpp} in which it is
possible to explain the number of generations and they are also very
predictive concerning new  theoretical ideas as extra
dimensions~\cite{extra} and the little Higgs
mechanism~\cite{lhiggs}. Those models also include an extra neutral
vector boson so that  there is, in general, a mixing of $Z$, the
vector boson of $SU(2)_L\subset SU(3)_L$, and $Z^\prime$ the gauge
boson related to the $SU(3)_L$ symmetry. Working in the $Z,Z^\prime$
basis [the parameterization in Eq.~(\ref{zgeral}) is also valid but
in these models there is no mixing in the kinetic term i. e.,
$\sin\xi=0$]  it means that the condition $\sin\phi\ll1$  can be
obtained  if in this case the energy scale $\Lambda\equiv v_\chi$,
with $v_\chi$ related to the $SU(3)_L$ symmetry,  is above the TeV
scale. Hence, it is usually believed that only approximately we can
have that $Z_1\approx Z$, even at the tree level. The same happens
with the neutral current couplings, $g^{i}_{V,A}$, which only
approximately coincide with $g^{i(SM)}_{V,A}$. This is true since
the corrections to the $Z$ mass and $g^{i}_{V,A}$ in these models,
assuming $v_\chi\gg v_W \simeq 246$ GeV,  are proportional to
$(v_W/v_\chi) ^{2}$ and for $v_\chi\to\infty$ we recover exactly the
SM with all its degrees of freedom, with the heavier ones introduced
by the $SU(3)_L$ symmetry decoupled. However, we expect that
$v_\chi$ should not be extremely large if new physics is predicted
to show up in the near future experiments. In practice, measurements
of the $\rho_0$ parameter, and FCNC processes like $\Delta M_K$,
should impose constraints upon the $v_\chi$ scale at which the
$SU(3)_L$ symmetry arises.

Let us consider, for instance,   the model of Ref.~\cite{331} in
which  the electric charge operator is defined as ${\mathcal
Q}=\left( T_3-\sqrt3T_8\right)+X$, where $T_i$ are the usual $SU(3)$
generators and $X$ the charge assigned to the abelian factor
$U(1)_X$.  Thus, the SM fermionic content is embedded in the
extended group according to the multiplets transforming under
$SU(3)_L$ $\otimes$ $U(1)_X$ as:  for leptons,
$\Psi_{aL}=(\nu_a,\,l^-_a,\,(l^-)^c_a)^T_L\sim({\bf3},0)$,
$\,a=e,\mu,\tau$ [the superscript $^c$ means charge conjugation
operation]; and for quarks, $Q_{mL}$ = $(d_m,\,u_m,j_m)^T_L$ $\sim$
$({\bf3}^*, -1/3);\;m=1,2$; $Q_{3L}$ = $(u_3,\,d_3,\,J)^T_L$ $\sim$
$({\bf3},2/3)$,  $u_{\alpha R}$ $\sim$ $({\bf1},2/3)$, $d_{\alpha
R}$ $\sim$ $({\bf1},-1/3),\,\alpha=1,2,3$, $j_{mR}$ $\sim$
$({\bf1},-4/3)$, and $J_R$ $\sim$ $({\bf1},5/3)$. Here  $j_m$ and
$J$ are new  quarks needed to complete the representations. To
generate masses for all these fields through spontaneous symmetry
breaking   three triplets of Higgs scalars   and a sextet  are
introduced; they are  $\eta$ = $(\eta^0,\,\eta^{-}_1,\,\eta^+_2)^T$
$\sim$ $(\textbf{3},0)$, $\rho$ = $(\rho^+,\,\rho^0,\,\rho^{++})^T$
$\sim$ $(\textbf{3},+1)$, $\chi$ = $(\chi^-,\,\chi^{--},\,\chi^0)^T$
$\sim$ $(\textbf{3},-1)$ and
\begin{equation}
S=\left(\begin{array}{lll}
\sigma^0_1 & h^-_1 & h^+_2\\
h^-_1 & H^{--}_1 & \sigma^0_2 \\
h^+_2 & \sigma^0_2 & H^{++}_2
\end{array}\right)\sim (\textbf{6},0).
\label{sextet}
\end{equation}
The   VEVs  in the neutral components  of the scalar multiplets are
defined as $\langle \eta^0_1\rangle=v_\eta/\sqrt2$,
$\langle\rho^0_1\rangle=v_\rho/\sqrt2$,
$\langle\chi^0_1\rangle=v_\chi/\sqrt2$ and
$\langle\sigma^0_2\rangle=v_s/\sqrt2$. It is also possible to have
$\langle\sigma^0_1\rangle\not=0$ giving Majorana mass to the
neutrinos, but we will not be concerned with this here.  The VEV
$\langle\chi^0_1\rangle$ reduces the symmetry to the SM
$SU(2)_L\otimes U(1)_Y$ symmetry and the other VEVs further reduce
it to the electromagnetic $U(1)_Q$ factor.

From the  kinetic terms for the scalar fields, constructed with the covariant derivatives
\begin{eqnarray}
{\mathcal D}_\mu\varphi=\partial_\mu\varphi - ig \vec W_\mu\cdot
\vec T \varphi  -ig_{_X} X\varphi B_\mu,\nonumber \label{dct}
\end{eqnarray}
\begin{equation}
{\mathcal D}_\mu S = \partial_\mu S -ig \left[ (\vec W_\mu\cdot \vec T)S +
S^T (\vec W_\mu\cdot \vec T)^T     \right] \label{dcs}
\end{equation}
where $g_{_X}$ denotes the U(1)$_X$ gauge coupling constant and
$\varphi=\eta,\rho,\chi$,  we obtain the mass matrices for the
vector bosons.  Besides $W^\pm$ there are two other charged vector
bosons, $V^\pm$ and $U^{\pm\pm}$. The masses of these charged vector
bosons are given exactly by  $M^2_W=(g^2/4)\,v^2_W$,
$M^2_V=(g^2/4)\,(v^2_\eta + 2v^2_s+v^2_\chi)$ and
$M^2_U=(g^2/4)\,(v_\rho^2 + 2v^2_s+v^2_\chi)$, where  we have
defined $v^2_{_W}=v_\eta^2 + v_\rho^2+2v^2_s$   [in models where
there are heavy leptons transforming non trivially under
$SU(3)_L\otimes U(1)_X$ there is not the contribution of the sextet
and the above equations are still valid simply doing
$v_s=0$~\cite{pt}].   For the mass square matrix of the neutral
vector bosons in this model we have the following form, after
defining  the dimensionless ratios $\overline{v}_\rho=
{v_\rho}/{v_\chi}$, $\overline{v}_{_W}={v_{_W}}/{v_\chi}$ and the
parameter $t^2= g_{_X}^2/g^2 = s_W^2/(1-4s_W^2)$,
\begin{eqnarray}
{\mathcal M}^2=\frac{g^2}{4}v_\chi^2\left(\begin{array}{ccc}
\overline{v}_{_W}^2 & \frac{1}{\sqrt 3}(
\overline{v}_{_W}^2-2\overline{v}_\rho^2 ) & -2t \overline{v}_\rho^2  \\
\frac{1}{\sqrt 3}( \overline{v}_{_W}^2-2\overline{v}_\rho^2 ) &
\frac{1}{3}(\overline{v}_{_W}^2 + 4)&
 \frac{2}{\sqrt 3}t(\overline{v}_\rho^2 +2) \\
-2t \overline{v}_\rho^2 & \frac{2}{\sqrt 3}t( \overline{v}_\rho^2 +2) &
4t^2(\overline{v}_\rho^2 +1)\end{array}\right),
\label{mnt}
\end{eqnarray}
in the $(W^3_\mu $, $W^8_\mu $,  $B_\mu)$ basis. This matrix has a
zero eigenvalue corresponding to the photon and two nonzero ones
which are given by
\begin{equation}
M_{Z_1}^2 = \frac{g^2v_\chi^2}{6}\left[3t^2( \overline{v}_\rho^2 +1
) +1+\overline{v}_{_W} ^2\right]  \left(1 - R \right),
\label{massntz}
\end{equation}
\begin{equation}
M_{Z_2}^2  =  \frac{g^2v_\chi^2}{6}\left[3t^2( \overline{v}_\rho^2
+1 ) +1+\overline{v}_{_W} ^2\right] \left(1 + R \right),
\label{massntzl}
\end{equation}
with
\begin{equation}
R=\left[1-\frac{3(4t^2+1)\left(  \overline{v}_{_W}
^2(\overline{v}_\rho^2 +1)-\overline{v}_\rho^4\right) }{\left(3t^2(
\overline{v}_\rho^2 +1 ) +1+ \overline{v}_{_W}
^2\right)^2}\right]^{\frac{1}{2}}.
\end{equation}

\textit{$\rho_1$ and $\rho_0$ parameters.}--- In order to analyze
the condition  which allows to identify $Z_1$ of the 3-3-1 model
with the $Z$ of the SM, let us introduce a dimensionless $\rho_1-$
parameter defined at the tree level as
$\rho_1=c^2_WM^2_{Z_1}/M^2_W$.  As we can see from Eqs.
(\ref{massntz}) and (\ref{massntzl}) both mass eigenvalues,
$M_{Z_1}$ and $M_{Z_2}$,  have a complicate dependence on the VEVs
but  we observe  from  Eq. (\ref{massntz})  that $\rho_1\leq1$ (or,
$M_{Z_1}\leq M_Z$) is a prediction of the model. Next, we can search
for the conditions under which we have $\rho_1\equiv\rho_0=1$, where
$\rho_0=c^2_WM^2_Z/M^2_W$ is the respective parameter in the SM.
This is equivalent to the condition that $M_{Z_1}\equiv M_Z$ at the
tree level. The equation  $\rho_1=1$ has besides the solution
$v_\chi\to\infty$, another less trivial one  which can be obtained
using Eq. (\ref{massntz}) above:
\begin{eqnarray}
\overline{v}^2_\rho= \frac{1-4s^2_W}{2c^2_W}\;\overline{v}^2_{_W}.
\label{sol3}
\end{eqnarray}
The condition in Eq.~(\ref{sol3}) implies, using the definition of $v_W$ given above
also $v^2_\eta+2v^2_s=[(1+2s^2_W)/2c^2_W]\;v^2_{_W}$. We recall that
the $U(1)_X$ quantum number of the $\eta$ and $S$ fields are different
from that of the $\rho$ field, so there is no symmetry among the
respective VEVs. We have verified that (\ref{sol3}) is stable in the following sense:
small deviations from it implies small deviations from $\rho_1=1$.
With $s^2_W=0.2312$~\cite{pdg} we obtain
$v_\rho\approx54$ GeV and $\sqrt{v^2_\eta+2v^2_s}\approx240$ GeV. Notice that
Eq.~(\ref{sol3}) is independent of the $v_\chi$ scale. Hence,
all consequences of it will be also independent of
$v_\chi$ as claimed above in  the $Introduction$. The fact that $v_\chi$ does not
need to have a large value to be consistent with the present phe\-no\-me\-no\-lo\-gy
is interesting in the models of Refs.~\cite{331,pt} since these models have a
Landau-like pole at the TeV scale~\cite{pl331}.

If we substitute Eq.~(\ref{sol3}) in the Eqs.~(\ref{massntz}) and
(\ref{massntzl}) we obtain $M^2_{Z_1}=(g^2/4c^2_W)\; v^2_{_W} \equiv
M_Z$ and
\begin{equation}
M^2_{Z_2}\equiv M^2_{Z^\prime}=\frac{g^2v^2_{_W} }{2}\frac{(1-2s^2_W)(4+
\overline{v}^2_{_W})+s^4_W(4-\overline{v}^4_{_W})}{6c^2_W(1-4s^2_W)}\;v^2_\chi.
\label{oba2}
\end{equation}
Thus, assuming that Eq.~(\ref{sol3}) is valid we shall not
distinguish between $Z_1$ and $Z$ and between $Z_2$ and $Z^\prime$
unless stated explicitly. Moreover, the mass of $Z^\prime$ can be
large even if $v_\chi$ is of the order of the electroweak scale. In
fact, from Eq.~(\ref{oba2}) we see that for $\overline{v}_{_W}=1$
(the electroweak scale is equal to the 3-3-1 scale) we obtain
$M_{Z^\prime}=3.77 M_W$. Of course for lower values of
$\overline{v}_{_W}$, $Z^\prime$ is heavier, for instance for
$\overline{v}_{_W}=0.25$ we have $M_{Z^\prime}=18.36 M_W$. We recall
that since $v_\chi$ does not contribute to the $W$ mass it is not
constrained by the 246 GeV upper bound. Thus, independently if
$\overline{v}^2_W$ is larger, smaller or equal to 1,  the charged
vector boson  $V$ is heavier than $U$,  being $\Delta
M=\sqrt{M_V^2-M_U^2}=75.96$ GeV, when $v_s=0$.

\textit{Neutral current couplings}.--- We have also obtained the
full analytical exact expressions for the neutral current couplings
$g^{i}_{V,A}$ and $f^{i}_{V,A}$, and verified that they also depend
on the VEVs in a complicated way. But when  Eq.~(\ref{sol3}) is used
in those expressions we obtain for the case of the known fermions
$g^{i}_{V,A}\equiv g^{i(SM)}_{V,A}$, and $f^{i}_{V,A}=
f^i_{V,A}(s_W)$, i. e., these couplings  depend only on the
electroweak mixing angle. For the lepton couplings with $Z^\prime$,
also after using Eq.~(\ref{sol3}) in the general expressions, we
obtain $f^\nu_V = f^\nu_A =f^l_V = -f^l_A=-
\sqrt{3(1-4s^2_W)}/6\,(\approx-0.07)$. We see that the couplings for
all leptons with $Z^\prime$ are leptophobic~\cite{dumm}.  In
particular,  the couplings of $Z$ to the exotic quarks $j_m$ and $J$
are given by $g^{j_m}_V = (8/3)s^2_W$, $g^J_V = -(10/3)s^2_W$ and
$g^{j_m}_A = g^J_A =0$. Notice also that the exotic quarks have pure
vectorial couplings with $Z$. The couplings of $Z^\prime$ in the
quark sector are given by:
\begin{eqnarray}
& &f^{u_m}_V = \frac{1}{2\sqrt{3}}\,\frac{1-6s^2_W}{\sqrt{1-4s^2_W}},
\; f^{u_m}_A =\frac{1}{2\sqrt{3}}\frac{1+2s^2_W}{\sqrt{1-4s^2_W}},
\nonumber \\
&&
f^{u_3}_V=-\frac{1}{2\sqrt{3}}\frac{1+4s^2_W}{\sqrt{1-4s^2_W}},\;
f^{u_3}_A=-\frac{1}{\sqrt3}\sqrt{1-4s^2_W},
\nonumber \\
&&
f^{d_m}_V=\frac{1}{2\sqrt{3}\sqrt{1-4s^2_W}},\;
f^{d_m}_A=\frac{\sqrt{1-4s^2_W}}{2\sqrt3}, \nonumber \\
&&
f^{d_3}_V=-\frac{1}{2\sqrt3}\frac{1-2s^2_W}{\sqrt{1-4s^2_W}} , \;
f^{d_3}_A = -\frac{1}{2\sqrt3}\frac{1+2s^2_W}{\sqrt{1-4s^2_W}},
 \nonumber \\
&& f^{j_m}_V =-\frac{1}{\sqrt3}\frac{1-9s^2_W}{\sqrt{1-4s^2_W}},\;
f^{j_m}_A =-\frac{1}{\sqrt3}\frac{c^2_W}{\sqrt{1-4s^2_W}},
\nonumber \\
&&
 f^J_V =\frac{1}{\sqrt3}\frac{1-11s^2_W}{\sqrt{1-4s^2_W}},\;
f^J_A =\frac{1}{\sqrt3}\frac{c^2_W}{\sqrt{1-4s^2_W}}.
\label{gvgaq2}
\end{eqnarray}
In literature~\cite{331,maperez} these couplings
were considered as an approximation of the exact couplings.
Notice that, all these couplings refer to fermions which are still
symmetry eigenstates, thus we see that in the leptonic sector there
are not FCNCs neither with $Z$ nor with $Z^\prime$ and, in the
quark sector there are FCNC only coupled to $Z^\prime$ as can be
see from Eq.~(\ref{gvgaq2}). However, FCNC mediated by the $Z^\prime$
depend only on its mass, but these FCNC are not necessary large since
there are also contributions in the scalar sector [see below].

The main feature introduced by the validity of the condition
Eq.~(\ref{sol3}), which we would like to stress, is the fact that
all couplings between the already known particles are exactly those
of the SM, regardless the value of the $v_\chi$ scale. Hence,
$v_\chi$ is not required to be large to recover those observed
couplings [until now $v_\chi \to \infty$ was the usual approach to
do that]. In this way, the 3-3-1 gauge symmetry could be realized,
for instance, at the electroweak scale ($v_\chi=v_W$) allowing the
extra particles introduced by the $SU(3)_L$ to be light enough to
not decouple and be discovered in the near future experiments.

\textit{A Goldberger-Treiman---like relation}.---
We can rewrite Eq.~(\ref{sol3}) as
\begin{equation}
g\;\frac{ v_\rho}{\sqrt2}= \frac{\sqrt{1-4s^2_W}}{c_W}\;M_W.
\label{sol3b}
\end{equation}
This is like the Goldberger-Treiman relation~\cite{gt58} in the sense that
its validity implies a larger symmetry of the model [see below] and all quantities
appearing in it can be measured independently of each other. In fact,
all but $v_\rho$, are already well known. However, cross sections of several
processes, for instance $e^+e^-\to ZH$ where $H$ is a neutral Higgs
scalar transforming as doublet of $SU(2)$, are sensitive to the value
of $v_\eta$ (or $v_\rho$)~\cite{cieza}. So, in principle it is possible to
verified if Eq.~(\ref{sol3}), or equivalently Eq.~(\ref{sol3b}), is
satisfied and if the 3-3-1 symmetry can be implemented near the weak scale.

\textit{Custodial symmetries and the oblique $T$ parameter}.---
We can understand the physical meaning of Eq.~(\ref{sol3}) in the following
way. The 3-3-1 models have an approximate $SU(2)$ custodial symmetry.
This is broken by the mixing between $Z$ and $Z^\prime$. In
general we have a mixture between these neutral bosons in such a way
that the mass eigenstates $Z_1$ and $Z_2$ can be written as~\cite{ng}
$Z_1=Zc_\phi-Z^\prime s_\phi$ and $Z_2= Zs_\phi+Z^\prime c_\phi$,
and the condition in Eq.~(\ref{sol3}) is equivalent to put $\phi=0$
i. e., no mixing at all between $Z$ and $Z^\prime$. There is also
an approximate $SU(3)$ custodial symmetry because when both (\ref{sol3})
and $\sin\theta_W=0$ are used, we have $M_U/M_{Z^\prime}=1$. However this
symmetry is badly broken. We stress that the alternative approach used in
literature~\cite{331,ng,cordero} is that the condition $\sin\phi<<1$ is
obtained by assuming that $v_\chi>>v_W$. This is of course still a
possibility if the relation (\ref{sol3}) is not confirmed experimentally.
However, we have shown above that it is possible that $\phi=0$
even if  $v_\chi=v_W$.

Of course, in any case radiative corrections will induce a mixing among
$Z$ and $Z^\prime$, i. e., a finite contribution to $\phi$. This should
imply small deviations from $\rho_0=1$.
The oblique $T$ parameter constraints this deviations since
$\rho_0-1\simeq \alpha T$ and it is given, for the 3-3-1 models,
in Ref.~\cite{inami}. Using the expressions of Ref.~\cite{inami}  but
without the mixing at the tree level ($\phi=0$ in Eq.~(4.1) of
\cite{inami}), we obtain for example $T=-0.1225$ for
$\overline{v}_{_W}=1$ and $T=-0.012$ for $\overline{v}_{_W}=0.25$,
with $T\to 0$ as $\overline{v}_{_W}\to0$ ($v_\chi\to\infty$), and all
$T$ values calculated with $\overline{v}_{_W}\leq 1$ are within the
allowed interval~\cite{pdg}. This implies that the condition
Eq.~(\ref{sol3}) is not significantly disturbed by radiative
corrections. It means that the natural value of $\sin\phi$, arisen
only through radiative corrections, is small because the symmetry of
the model is augmented when this parameter vanishes.

The important thing is that even if Eq.~(\ref{sol3}) or equivalently
Eq.~(\ref{sol3b}) are valid only approximately, we will have that again
$M_{Z_1}\approx M_Z$ and also the neutral current couplings of $Z_1$ only
approximately coincide with those of the SM, but now this
is valid almost independently of the value of $v_\chi$. That is, the
3-3-1 symmetry still can be implemented at an energy scale near the
electroweak scale.

\textit{Experimental constraints on the $SU(3)_L$ scale}.--- Once
the $v_\chi$ scale is arbitrary when Eq.~(\ref{sol3}) is satisfied,
we can ask ourselves what about the experimental limit upon the
masses of the extra particles that appear in  the model. After all
they depend mainly on $v_\chi$, the scale at which the $SU(3)_L$
symmetry is supposed to be valid. Firstly, let us consider the
$Z^\prime$ vector boson. It contributes to the $\Delta M_K$ at the
tree level~\cite{dumm94}. This parameter imposes constraints over
the quantity $({\mathcal  O}^d_L)_{3d}({\mathcal
O}^d_L)_{3s}\,(M_Z/M_{Z^\prime})$, which must be of the order of
$10^{-4}$ to have compatibility with the measured $\Delta M_K$. This
can be achieved with $M_{Z^\prime}\sim 4$ TeV if we assume that the
mixing matrix have a Fritzsch-structure ${\mathcal
O}^d_{Lij}=\sqrt{m_j/m_i}$~\cite{sher} or, it is possible that the
product of the mixing angles saturates the value
$10^{-4}$~\cite{dumm94}, in this case $Z^\prime$ can have a mass
near the electroweak scale. More important  is the fact that there
are also in this model FCNC mediated by the neutral Higgs scalar
which contributes to $\Delta M_K$. These contributions depend on the
mixing matrix in the right-hand $d$-quark sector, ${\mathcal
O}^d_R$, and also on some Yukawa couplings, $\Gamma^d$, i. e., the
interactions are of the form $\bar{d}_L({\mathcal
O}^d_L)_{d3}\Gamma^d_{3\alpha} ({\mathcal  O}_R)_{\alpha}s_R$. Thus,
their contributions to $\Delta M_K$ may have opposite sing relative
to that of the contribution of $Z^\prime$. A realistic calculation
of the $\Delta M_K$ in the context of 3-3-1 models has to take into
account these extra contributions as well. Muonium-antimuonium
transitions would imply a lower bound of 850 GeV on the masses of
the doubly charged gauge bileptons, $U^{--}$~\cite{willmann99}.
However this bound depends on assumptions on the mixing matrix in
the lepton charged currents coupled to $U^{--}$ and also it does not
take into account that there are in the model doubly charged scalar
bileptons which also contribute to that transition~\cite{pleitez00}.
Concerning these doubly charged scalars, the lower limit for their
masses are only of the order of 100 GeV~\cite{d0}. From fermion pair
production at LEP and lepton flavor violating effects suggest a
lower bound of 750 GeV for the mass $M_U$ but again it depends on
assumptions on the mixing matrix and on the assumption that those
processes are induced only by the $U^{--}$ boson~\cite{tully}. Other
phenomenological analysis in $e^+e^-,e\gamma$ and $\gamma\gamma$
colliders assume bileptons with masses between 500 GeV and 1
TeV~\cite{dion1,dion2}. The muonium fine structure only implies
$M_U/g>215$ GeV~\cite{fujii} but also ignores the contributions of
the doubly charged scalars. Concerning the exotic quark masses there
is no lower limit for them but if they are in the range of 200-600
GeV they may be discovered at the LHC~\cite{yara}. Direct search for
quarks with $Q=(4/3)$e imply that they are excluded if their mass is
in the interval 50-140 GeV~\cite{abe} but only if they are stable.
Similarly, most of the searches for extra neutral gauge bosons are
based on models that do not have the couplings with the known
leptons and quarks as those of the 3-3-1 model~\cite{babu}, anyway
we have seen that even if $v_\chi=v_W$ the $Z^\prime$ has a mass of
the order of 300 GeV. Finally, rare processes like $\mu^-\to
e^-\nu_e\nu^c_\mu$, induced by the extra particles are also not much
restrictive. We may conclude that there are not yet definitive
experimental bounds on the masses of the extra degrees of freedom of
the 3-3-1 models.

\textit{Conclusions}.--- Summarizing, we have shown that  if the
condition in Eq.~(\ref{sol3}) is realized, concerning the already
known particles, the 3-3-1 model and the SM are indistinguishable
from each other at the tree level and, as suggested by the value of
the $T-$parameter obtained above, it is possible that this happens
even at the one loop level. The models can be confirmed or  ruled
out by searching directly for the effects of their new particles,
for instance in left-right asymmetries in lepton-lepton scattering.
An  asymmetry of this sort only recently has begun to be measured in
an electron-electron fixed target experiment~\cite{e158}, but the
effects of these asymmetries could be more evident in collider
experiments~\cite{assi,yara2}. From all we have discussed above, it
is clear that new physics may be really just around the corner.

We have also verified that in 3-3-1 models with heavy
leptons~\cite{pt} and with  right-handed neutrinos transforming
non-trivially under the 3-3-1 gauge symmetry~\cite{mpp} a similar
situation occurs, but, in the later model, the equivalent of the
relation in Eq~(\ref{sol3}) is given by
$\overline{v}^2_\rho=[(1-2s^2_W)/2c^2_W]\overline{v}^2_{W}$~\cite{dong}.
Notwithstanding we have verified that the latter condition is less
stable, in the sense we said before: small deviation of it implies
large deviation from $\rho_1=1$. This suggests that, when both 3-3-1
models were embedding in a $SU(4)_L\otimes U(1)_N$ model~\cite{341},
the $SU(3)$ subgroup which contains  the vector boson $Z$ of the SM
should be the minimal 3-3-1 model considered in this work.

A. G. D. is supported by FAPESP under the process 05/52006-6, and this work
was partially supported by CNPq under the processes 305185/03-9 (JCM), and
306087/88-0~(VP).


\begin{thebibliography}{99}
\bibitem{lr} J. C. Pati, A. Salam, Phys. Rev. D \textbf{10} (1974) 275;
R. N. Mohapatra, J. C. Pati, \textit{ibid}, D \textbf{11} (1975) 2558; G. Senjanovic,
Nucl. Phys. \textbf{B153} (1979) 334.
\bibitem{hewett} J. Hewett, T. Rizzo, Phys. Rep. {\bf183} (1989) 193; F.
del Aguila, Acta Phys. Polo. \textbf{B25} 1317 (1994); A. Leike, Phys. Rep.
\textbf{317} (1999) 143.
\bibitem{lh} N. Arkani-Hamed, A. G. Cohen, E. Katz,  A. E. Nelson, JEHP
\textbf{0207} (2002) 034.
\bibitem{kk} I. Antoniadis, K. Benakli, Int. J. Mod. Phys. \textbf{A15} (2000)  4237.
\bibitem{lhcilc} G. Weiglein \textit{et al}. (LHC/ILC Study Group), hep-ph/0410364.
\bibitem{babu} K. S. Babu, in Ref.~\cite{pdg} p. 380 and references therein.
\bibitem{pdg} S. Eidelman, \textit{et al.} (PDG Collaboration), Phys. Lett.
\textbf{B592} (2004) 1.
\bibitem{carena} M. Carena, A. Daleo, B. A. Dobrescu, T. M. P. Tait, Phys.
Rev. D {\bf70} (2004) 093009; J. Kang, P. Langacker, Phys. Rev. D {\bf71} (2005) 035014.
\bibitem{331} F. Pisano, V. Pleitez, Phys. Rev. D \textbf{46} (1992) 410;
R. Foot, O. F. Hernandez, F. Pisano, V. Pleitez, Phys. Rev. D \textbf{47} (1993)
4158; P. \ Frampton, Phys.\ Rev.\ Lett.\ \textbf{69} (1992) 2889.
\bibitem{pt} V. Pleitez, M. D. Tonasse, Phys. Rev. D \textbf{48} (1993) 2353.
\bibitem{mpp} J. C. Montero, F. Pisano,  V. Pleitez, Phys. Rev. D
\textbf{47} (1993) 2918; H. N. Long, \textit{ibid}. D \textbf{54}(1996) 4691.
\bibitem{extra} S. Dimopoulos, D. E. Kaplan, Phys. Lett. \textbf{B531} (2002)
127; W.-F. Chang, J. N. Ng, Phys. Rev. D \textbf{69} (2004) 056005.
\bibitem{lhiggs} M. Schmaltz, JHEP \textbf{0408} (2004) 056; T. Roy, M. Schmaltz,
\textit{ibid}. \textbf{0601} (2006) 149, hep-ph/0509357.
\bibitem{pl331} A. G. Dias, R. Martinez, V. Pleitez, Eur. Phys. J
\textbf{C39} (2005) 101; A. G. Dias, Phys. Rev. D \textbf{71} (2005) 015009.
\bibitem{dumm} D. G\'omez Dumm, Phys. Lett. \textbf{B411} (1997) 313.
\bibitem{maperez} M. A. P\'erez, G. Tavares-Velasco,  J. J. Toscano, Phys.Rev. D
\textbf{69} (2004) 115004.
\bibitem{gt58} M. L. Goldberger, S. B. Treiman, Phys. Rev. \textbf{110} (1958)
1178, 1478.
\bibitem{cieza} J. E. Cieza-Montalvo, M. D. Tonasse, Phys. Rev. D \textbf{71}
(2005) 095015.
\bibitem{ng} D. Ng. Phys. Rev. D \textbf{49} (1994) 4805.
\bibitem{cordero} A. Cordero-Cid, G. Tavares-Velasco, J. J. Toscano, Phys. Rev. D
\textbf{72} (2005)  057701.
\bibitem{inami} H. N. Long, T. Inami, Phys. Rev. D \textbf{61} (2000)  075002.
\bibitem{dumm94} D. G. Dumm, F. Pisano, V. Pleitez, Int. J. Mod. Phys.
\textbf{A9} 1609 (1994); J. T. Liu, Phys. Rev. D \textbf{50} (1994) 542.
\bibitem{sher} J.-Alexis Rodriguez, M. Sher, Phys. Rev. D \textbf{70} (2004)
117702.
\bibitem{willmann99} L. Willmann {\it et. al.}, Phys. Rev. Lett. \textbf{82} (1999)
49; B. L. Robert, M. Grassi,  A. Sato, hep-ex/0510055.
\bibitem{pleitez00} V. Pleitez, Phys. Rev. D \textbf{61} (2000) 057903.
\bibitem{d0} V. M. Abazov, \textit{et al.} (D0 Collaboration), Phys. Rev. Lett. \textbf{93}
(2004) 141801.
\bibitem{tully} M. B. Tully, G. C. Joshi, Phys. Lett. \textbf{B466} (1993)  333;
M. Raidal, Phys. Rev. D \textbf{57}  (1998) 2013.
\bibitem{dion1} N. Lepor\'e, B. Thorndyke, H. Nadeau, D. London, Phys. Rev. D
\textbf{50} (1994) 2031.
\bibitem{dion2}  K. Sasaki, K. Tokushuku, S. Yamada, Y. Yamazaki, Phys.
Lett.~\textbf{ B345}  (1995) 495; B. Dion, T. Gregoire, D. London, L. Marleau, H. Nadeau,
Phys. Rev. D \textbf{59}  (1999) 075006.
\bibitem{fujii} H. Fujii, Y. Mimura, K. Sasaki, T. Sasaki, Phys. Rev. D \textbf{49} (1994) 559.
\bibitem{yara} Y. A. Coutinho, P. P. Queiroz Filho, M. D. Tonasse, Phys. Rev. D
\textbf{60} (1999) 115001.
\bibitem{abe} P. Abe \textit{et al.} (CDF Collaboration), Phys. Rev. D
\textbf{46} (1992) 1889.
\bibitem{e158} P. L. Anthony {\it et al.} (E158 Collaboration), Phys. Rev. Lett. {\bf92} (2004) 181602.
\bibitem{assi} J. C. Montero, V. Pleitez, M. C. Rodriguez, Phys. Rev. D \textbf{58} (1998) 094026;
\textit{ibid}. \textbf{58} (1998) 097505;  Int. J. Mod. Phys. {\bf A16} (2001) 1147--1160;
hep-ph/9903317; hep-ph/0204130 and references therein.
\bibitem{yara2} E. Ramirez, Y. A. Coutinho,  J. S\'a Borges, Phys. Lett. \textbf{B632}  (2006) 675.
\bibitem{dong}  P. V. Dong, H. N. Long, Eur. Phys. J. \textbf{C 42} (2005) 325.

\bibitem{341} F. Pisano, V. Pleitez, Phys. Rev. D {\bf51} (1995) 3865.
\end{thebibliography}
\end{document}